\documentclass[10pt, technote]{IEEEtran}
\ifCLASSINFOpdf
\else
\fi
\usepackage[pdftex]{graphicx}
\graphicspath{{../pdf/}{../jpeg/}}
\DeclareGraphicsExtensions{.pdf,.jpeg,.png}

\usepackage{amsmath}
\ifCLASSOPTIONcompsoc
  \usepackage[caption=false,font=normalsize,labelfont=sf,textfont=sf]{subfig}
\else
  \usepackage[caption=false,font=footnotesize]{subfig}
\fi
\usepackage{color}

\begin{document}
%
\title{Non-singular field-only surface integral equations for electromagnetic scattering}
%
%
%

\author{Evert~Klaseboer, Qiang~Sun and~Derek~Y.~C.~Chan
\thanks{E. Klaseboer, Institute of High Performance Computing, 1 Fusionopolis Way, 138632, Singapore. Email: Evert@ihpc.a-star.edu.sg}
\thanks{Q. Sun (Corresponding Author), Department of Chemical and Biomolecular Engineering, University of Melbourne, Parkville, VIC, 3010 Australia. Email: Qiang.Sun@unimelb.edu.au}
\thanks{D. Y C. Chan, School of Mathematics and Statistics, University of Melbourne, Parkville, VIC, 3010 Australia, and Department of Chemistry and Biotechnology, Swinburne University of Technology, Hawthorn, VIC, 3122, Australia. Email: D.Chan@unimelb.edu.au}
\thanks{Manuscript received Sep 29, 2016.}}

%
%

\markboth{IEEE Transactions on Antennas and Propagation,~Vol.~65, No.~2, PP.~972~(2017),~DOI:~10.1109/TAP.2016.2632619}%
{Klaseboer \MakeLowercase{\textit{et al.}}: No-singular field-only surface integal equations}
%



\maketitle

\begin{abstract}
A boundary integral formulation of electromagnetics that involves only the components of $\boldsymbol{E}$ and $\boldsymbol{H}$ is derived without the use of surface currents that appear in the classical PMCHWT formulation. The kernels of the boundary integral equations for $\boldsymbol{E}$ and $\boldsymbol{H}$ are non-singular so that all field quantities at the surface can be determined to high precision and also geometries with closely spaced surfaces present no numerical difficulties. Quadratic elements can readily be used to represent the surfaces so that the surface integrals can be calculated to higher numerical precision than using planar elements for the same numbers of degrees of freedom.
\end{abstract}


%
\IEEEpeerreviewmaketitle

\section{Introduction}

\IEEEPARstart{T}{he} surface integral or boundary integral formulation of frequency domain electromagnetics was established by the classic works of Poggio \& Miller~\cite{PoggioMiller_1973}, Chang \& Harrington~\cite{ChangHarrington_1977} and Wu \& Tsai~\cite{WuTsai_1977} (PMCHWT) over 40 years ago and has been widely used ever since. In the PMCHWT formulation, the electric and magnetic fields, $\boldsymbol{E}$ and $\boldsymbol{H}$, are given in terms of electric and magnetic surface currents or equivalently scalar and vector potentials~\cite{Harrington_1989} that are found by solving surface integral equations. The fields $\boldsymbol{E}$ and $\boldsymbol{H}$ are then obtained by post-processing the surface current values. Many numerical methods have been developed to solve the surface current integral equations. A popular scheme is to use the Rao-Wilton-Glisson (RWG)~\cite{RaoWiltonGlisson_1982} basis functions that enforce charge conservation to represent the surface currents on planar triangular surface elements. It has been pointed out recently that this development is still not without challenges~\cite{Wu_2014}. The evaluation of field quantities gives rise to integral equations with hypersingular kernels due to dyadic Green's functions~\cite{Chao_1995} that introduce additional numerical difficulties in the zero frequency or long wavelength limit~\cite{Chew_2009}.

The well-known analytical solution of the scattering of an electromagnetic plane wave by a single sphere uses two scalar Debye potentials that satisfy the scalar Helmholtz equation~\cite{Wilcox_1956, vdHulst_1957, Liou_1977}. Here, motivated by the conciseness of this approach, we develop a fundamental reformulation of electromagnetics that works directly with field variables that satisfy scalar Helmholtz equations. In contrast to the PMCHWT approach, it is not necessary to solve for surface current densities as intermediate quantities. The scalar Helmholtz equations for the field components are solved by a recently developed boundary integral method in which all surface integrals have singularity-free integrands and the term involving the solid angle is eliminated~\cite{Klaseboer_2012, Sun_2015}. Thus the consequential advantages are: (i)~components of $\boldsymbol{E}$ and $\boldsymbol{H}$ are computed directly; (ii)~field quantities at or near surfaces can be calculated without loss of precision; (iii)~geometries where different parts of surfaces are very close together do not have detrimental effects on the numerical accuracy; (iv)~the ease with which higher order surface elements can be  used to represent boundaries more accurately enables the boundary integrals to be evaluated using standard quadrature and yet confer high numerical accuracy with fewer degrees of freedom and (v)~the accuracy of the numerical implementation means that the effect of any resonant solutions of the Helmholtz equation are negligible unless the wavenumber is extremely close to the resonant values, so that the resonant solution should not affect practical applications if the present approach is used.

\section{Theory}

We illustrate our electromagnetics formulation with the scattering problem by 3D perfect electrical conductors (PEC). The  generalisation to dielectric scatterers involves more complex algebra, yet is based on the same physical concepts \cite{Sun_2016b}. In the frequency domain with time dependence $\exp(j \omega t)$, the propagating electric field $\boldsymbol{E}$ in a source free region is given by the wave equation ($k^2 \equiv \omega^2 \epsilon_r \epsilon_0 \mu_r \mu_0 \equiv \omega^2 \epsilon \mu$): 
\begin{eqnarray} \label{eq:Maxwell}
  \nabla^2 \boldsymbol{E} + k^2 \boldsymbol{E} = \boldsymbol{0} \qquad \text{with}  \qquad \nabla \cdot \boldsymbol{E} = 0.
\end{eqnarray} 
Since $\nabla \cdot \boldsymbol{E} = 0$, there are only two independent components of $\boldsymbol{E}$ in~(\ref{eq:Maxwell}) and they are found by specifying the incident field, $\boldsymbol{E}^i = \boldsymbol{E}_0 \exp(-j \boldsymbol{k \cdot r)}$, where $\boldsymbol{r}=(x,y,z)$ is the position vector, and imposing the boundary condition that the tangential components of $\boldsymbol{E}$ must vanish on the surface, $S$ of the PEC.

The condition $\nabla \cdot \boldsymbol{E} = 0$ can be replaced using a vector identity for ($\boldsymbol{r \cdot E}$) to give 
\begin{eqnarray}
  \nabla^2 \boldsymbol{E} + k^2 \boldsymbol{E} = \boldsymbol{0}  \qquad  \qquad \label{eq:Helm_E} \\ 
  2 (\nabla \cdot \boldsymbol{E}) \equiv \nabla^2 (\boldsymbol{r \cdot E}) + k^2 (\boldsymbol{r \cdot E}) = 0. \label{eq:Helm_xE}
\end{eqnarray} 
The results in (\ref{eq:Helm_E}) and (\ref{eq:Helm_xE}) were first demonstrated explicitly by Lamb for elastic vibrations~\cite{Lamb_1881}. They are independent of the choice of the origin of the coordinate system as can be verified by adding a constant vector to $\boldsymbol{r}$. However, they have significant relevance to electromagnetics in that they show $\boldsymbol{E}$ is determined directly by a coupled set of 4 scalar Helmholtz equations: 
\begin{eqnarray} \label{eq:Helm_p}
  \nabla^2 p_i(\boldsymbol{r}) + k^2 p_i(\boldsymbol{r}) = 0,  \qquad i = 1 .. 4
 \end{eqnarray} 
that we will solve by the boundary integral method. The scalar functions $p_i(\boldsymbol{r})$ denote one of the 3 Cartesian components of $\boldsymbol{E}$ or $(\boldsymbol{r \cdot E}$). Equation (\ref{eq:Helm_E}) furnishes 3 relations between the 6 unknowns: $E_{\alpha}$ and $\partial E_{\alpha} / \partial n$, $(\alpha =x,y,z)$, where $\partial/\partial n \equiv \boldsymbol{n} \cdot \nabla$ and $ \boldsymbol{n}$ is the outward unit normal of the surface, $S$ of the solution domain. Equation (\ref{eq:Helm_xE}) between $(\boldsymbol{r} \cdot \boldsymbol{E})$ and $\partial (\boldsymbol{r} \cdot \boldsymbol{E}) / \partial n$ provides one more relation between $E_{\alpha}$ and $\partial E_{\alpha} / \partial n$ since: $\partial (\boldsymbol{r} \cdot \boldsymbol{E}) / \partial n = \boldsymbol{n} \cdot \boldsymbol{E} + \boldsymbol{r} \cdot \partial \boldsymbol{E} / \partial n$. The electromagnetic boundary conditions on the continuity of the tangential components of $\boldsymbol{E}$ provide the remaining 2 equations to determine $\boldsymbol{E}$ and $\partial \boldsymbol{E} / \partial n$ completely.

For scattering by a PEC, it is more convenient to work in terms of the normal component, $E_n = \boldsymbol{n} \cdot \boldsymbol{E}$, and two tangential components, $\boldsymbol{E}_t = (E_{t1}, E_{t2}) $, of the electric field at the surface. In the PEC case, there are 4 unknowns to be determined, namely: $\partial E_x/ \partial n, \partial E_y/ \partial n, \partial E_z/ \partial n, E_n$ because the tangential components of the electric field must vanish on the surface of a PEC. We decompose $\boldsymbol{E}$ into a sum of the incident field, $\boldsymbol{E}^{i}$ and the scattered field, $\boldsymbol{E}^{s}$ so on the surface of the PEC, the tangential components of the scattered field cancel those of the incident field. Physically, $E_n$ is proportional to the induced surface charge density on the PEC. Thus the number of unknowns to be found is the same as for the classic solution of the scattering problem by a PEC sphere using a pair of scalar Debye potentials in which the 2 unknown functions and their derivatives have to be found~\cite{vdHulst_1957, Liou_1977}. However, in the Debye potential approach, the electromagnetic boundary conditions are expressed as combinations of the two potentials and components of their gradients on the surface of the PEC and give rise to equations that are not straightforward to solve in the framework of the boundary integral method.

The boundary integral solution of~(\ref{eq:Helm_p}) for the scattered field is based on Green's Second Identity that gives a relation between $p_i(\boldsymbol{r})$ and its normal derivative $\partial{p_i}/\partial{n}$ at points $\boldsymbol{r}$ and $\boldsymbol{r}_0$ on the boundary, $S$.  All singularities associated with the Green's function $G \equiv G(\boldsymbol{r},\boldsymbol{r}_0) = \exp(-jk|\boldsymbol{r} - \boldsymbol{r}_0|) / |\boldsymbol{r} - \boldsymbol{r}_0|$, can be removed analytically to give~\cite{Klaseboer_2012, Sun_2015}
\begin{eqnarray} \label{eq:NS_BIE}
\int_{S}^{} {[p_i (\boldsymbol{r}) - p_i (\boldsymbol{r}_0) g(\boldsymbol{r}) - \frac{\partial {p_i(\boldsymbol{r}_0)}} {{\partial {n}}}  f(\boldsymbol{r})] \frac{\partial {G}} {{\partial {n}}} dS(\boldsymbol{r}}) = \qquad \nonumber \\  \int_{S}^{} {G [ \frac{\partial {p_i (\boldsymbol{r})}} {{\partial {n}}} - p_i (\boldsymbol{r}_0) \frac{\partial {g (\boldsymbol{r})}} {{\partial {n}}} - \frac{\partial {p_i(\boldsymbol{r}_0)}} {{\partial {n}}} \frac{\partial {f (\boldsymbol{r})}} {{\partial {n}}} ]dS(\boldsymbol{r}}). 
\end{eqnarray} 
The requirement on $f(\boldsymbol{r})$ and $g(\boldsymbol{r})$ is that they satisfy the Helmholtz equation and the following conditions at $\boldsymbol{r} = \boldsymbol{r}_0$ on surface, $S$: $f(\boldsymbol{r}) = 0,\boldsymbol{n} \cdot  \nabla f(\boldsymbol{r}) = 1, g(\boldsymbol{r}) = 1, \boldsymbol{n} \cdot \nabla g(\boldsymbol{r}) = 0$. Examples of  possible choices of $f(\boldsymbol{r})$ and $g(\boldsymbol{r})$ can be found in~\cite{Klaseboer_2012,Sun_2015}. Thus if $p_i$ (or $\partial{p_i}/\partial{n}$) is given, then~(\ref{eq:NS_BIE}) can be solved for $\partial{p_i}/\partial{n}$ (or $p_i$) in a straightforward manner. The reason is that for $f(\boldsymbol{r})$ and $g(\boldsymbol{r})$ that obey the above conditions, the terms that multiply $G$ and $\partial G/\partial n$ vanish at the same rate as the rate of divergence of $G$ or $\partial G/\partial n$ as $\boldsymbol{r} \rightarrow \boldsymbol{r}_0$ and consequently both integrals have non-singular integrands and can thus be evaluated accurately by quadrature, see~\cite{Klaseboer_2012, Sun_2015} for details. Note that the solid angle at $\boldsymbol{r}_0$ has also been eliminated in (\ref{eq:NS_BIE}).

With the removal of all singular behavior and without the need to represent surface current densities, quadratic surface elements can be used to represent the surface geometry more accurately. This can provide orders of magnitude improvement in the numerical integration over standard methods (with singular integrands) for the same number of degrees of freedom~\cite{Sun_2015}. Once the field quantities are known on the boundary, values in the 3D solution domain, even at locations close to the boundaries can be obtained easily and accurately since the boundary integral equations are not singular~\cite{Klaseboer_2012,Sun_2015}. 

The formulation for the magnetic field, $\boldsymbol{H}$ is similar:
\begin{eqnarray} \label{eq:OurMaxwell_H}
  \nabla^2 \boldsymbol{H} + k^2 \boldsymbol{H} = \boldsymbol{0}  \qquad  \qquad  \label{eq:Helm_H} \\ 
  2 (\nabla \cdot \boldsymbol{H}) \equiv \nabla^2 (\boldsymbol{r \cdot H}) + k^2 (\boldsymbol{r \cdot H}) = 0 \label{eq:Helm_xH}
\end{eqnarray} 
but at PEC boundaries, (\ref{eq:Helm_xH}) is equivalent to the simpler condition that the normal component of $\boldsymbol{H}$ vanishes on the PEC:
\begin{eqnarray} \label{eq:Hn_eq_0}
     \boldsymbol{n \cdot H} = 0 \quad \text{on} \quad S.
\end{eqnarray}
To apply the boundary condition on the tangential components of $\boldsymbol{E}$, we choose two orthogonal unit tangents $\boldsymbol{p}$ and $\boldsymbol{t}$ on $S$, and use Ampere's law to express the component of  $\boldsymbol{E}$ parallel to $\boldsymbol{p}$, namely, $E_p \equiv \boldsymbol{E} \boldsymbol{\cdot} \boldsymbol{p} = \boldsymbol{E} \cdot (\boldsymbol{t} \times \boldsymbol{n})$, in terms of $\boldsymbol{H}$
\begin{eqnarray} 
  E_p &=& \boldsymbol{t} \cdot (\boldsymbol{n} \times \boldsymbol{E})  = \frac{1}{j \omega \epsilon} \{ \boldsymbol{t} \cdot (\boldsymbol{n} \times \nabla \times \boldsymbol{H}) \} \label{eq:Et} \nonumber \\
         &=& \frac{1}{j \omega \epsilon} \{ \boldsymbol{n} \cdot (\boldsymbol{t} \cdot \nabla) \boldsymbol{H} - \boldsymbol{t} \cdot (\boldsymbol{n} \cdot \nabla) \boldsymbol{H} \}  = 0. \label{eq:Et_EQ_0} 
\end{eqnarray}
The second equality in (\ref{eq:Et_EQ_0}) follows from the electric field boundary condition on the PEC surface, $S$. 

Our formulation for PEC problems for $\boldsymbol{H}$, in (\ref{eq:Helm_H}) - (\ref{eq:Et_EQ_0}), is slightly more complex than our formulation for $\boldsymbol{E}$, in (\ref{eq:Helm_E}) - (\ref{eq:Helm_xE}), because of the need to use (\ref{eq:Et_EQ_0}) to impose the PEC boundary condition for $\boldsymbol{E}$ in terms of $\boldsymbol{H}$. 

\section{Numerical implementation}

\begin{figure}[!t]
\centering{}\includegraphics[width=3.4in]{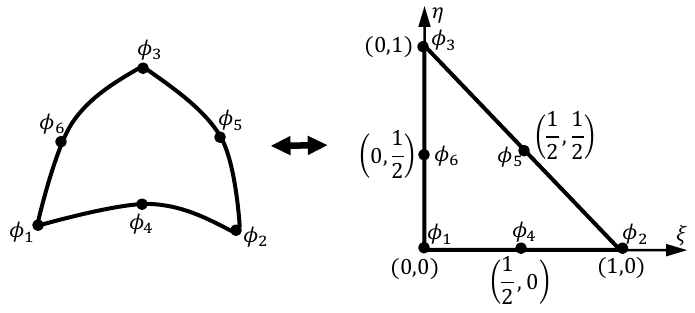} \caption{\label{Fig:quadratic} 
The  interpolation scheme on a quadratic surface element in the local surface variables ($\xi$, $\eta$).} 
\end{figure}

\begin{figure*} [!t]
  \centering{}
  \subfloat[]{ \includegraphics[width=3.3in]{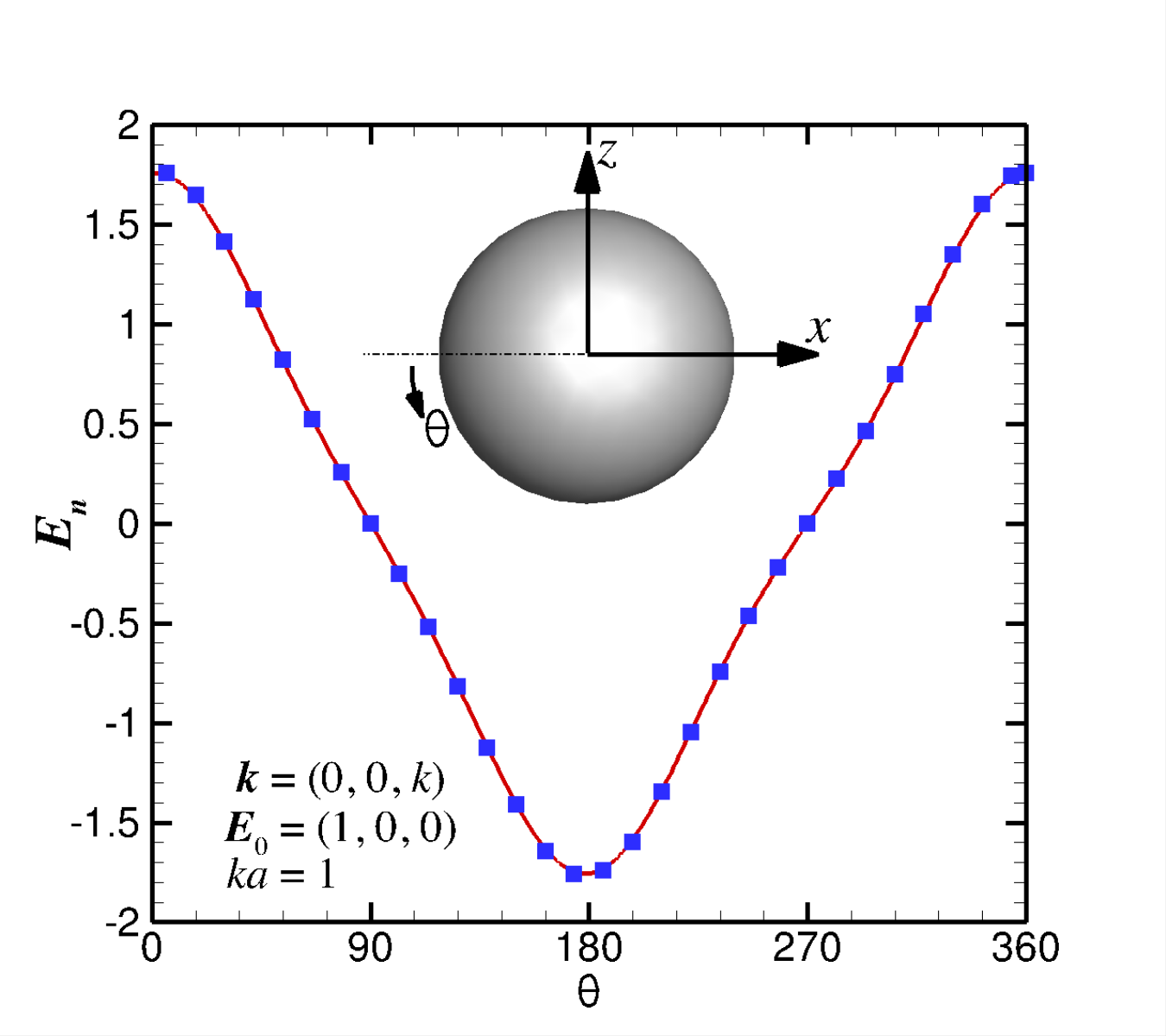} }
  \subfloat[]{ \includegraphics[width=3.3in]{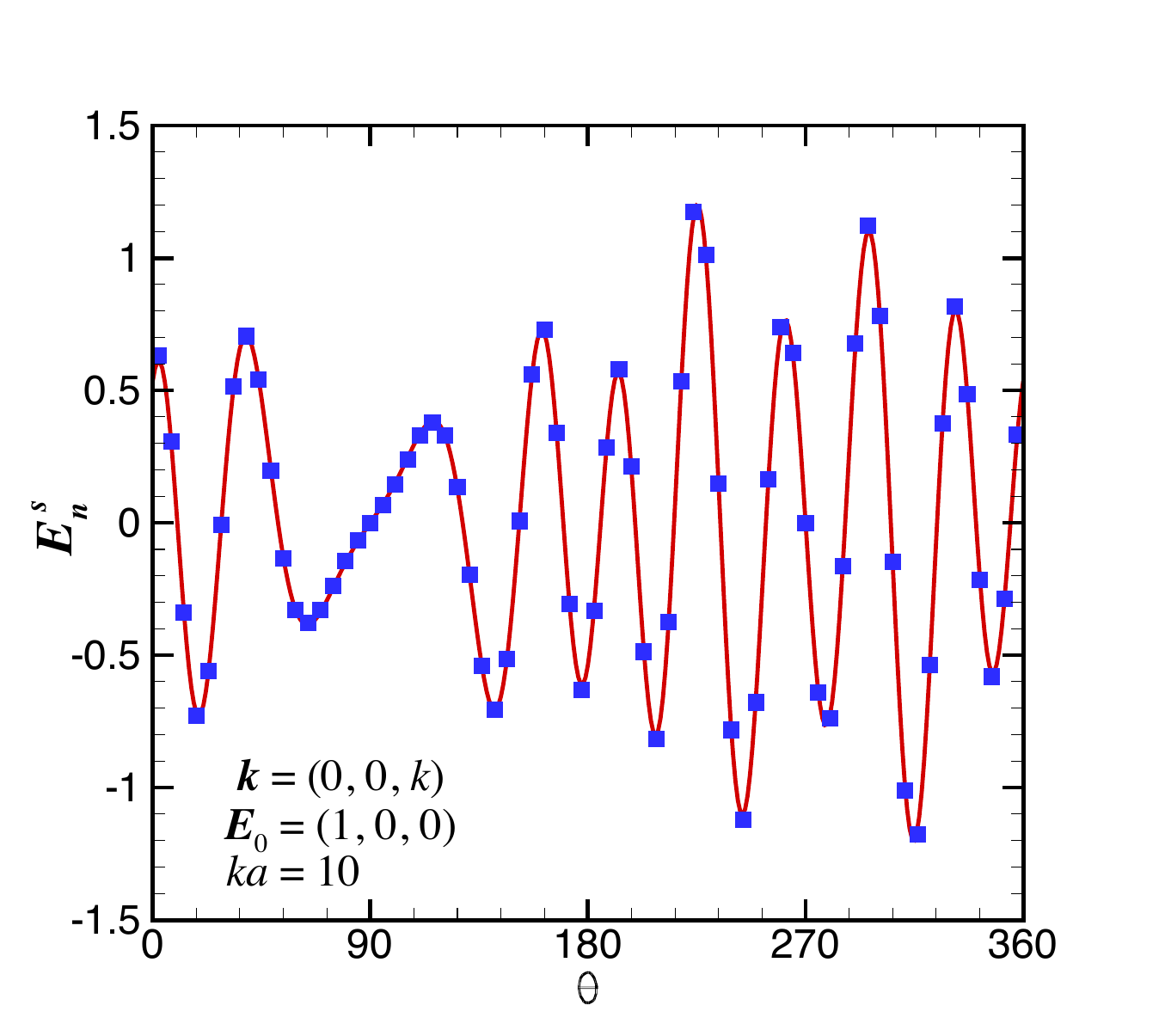} } \\
  \subfloat[$ka=10, x=0$]{ \includegraphics[width=3.3in]{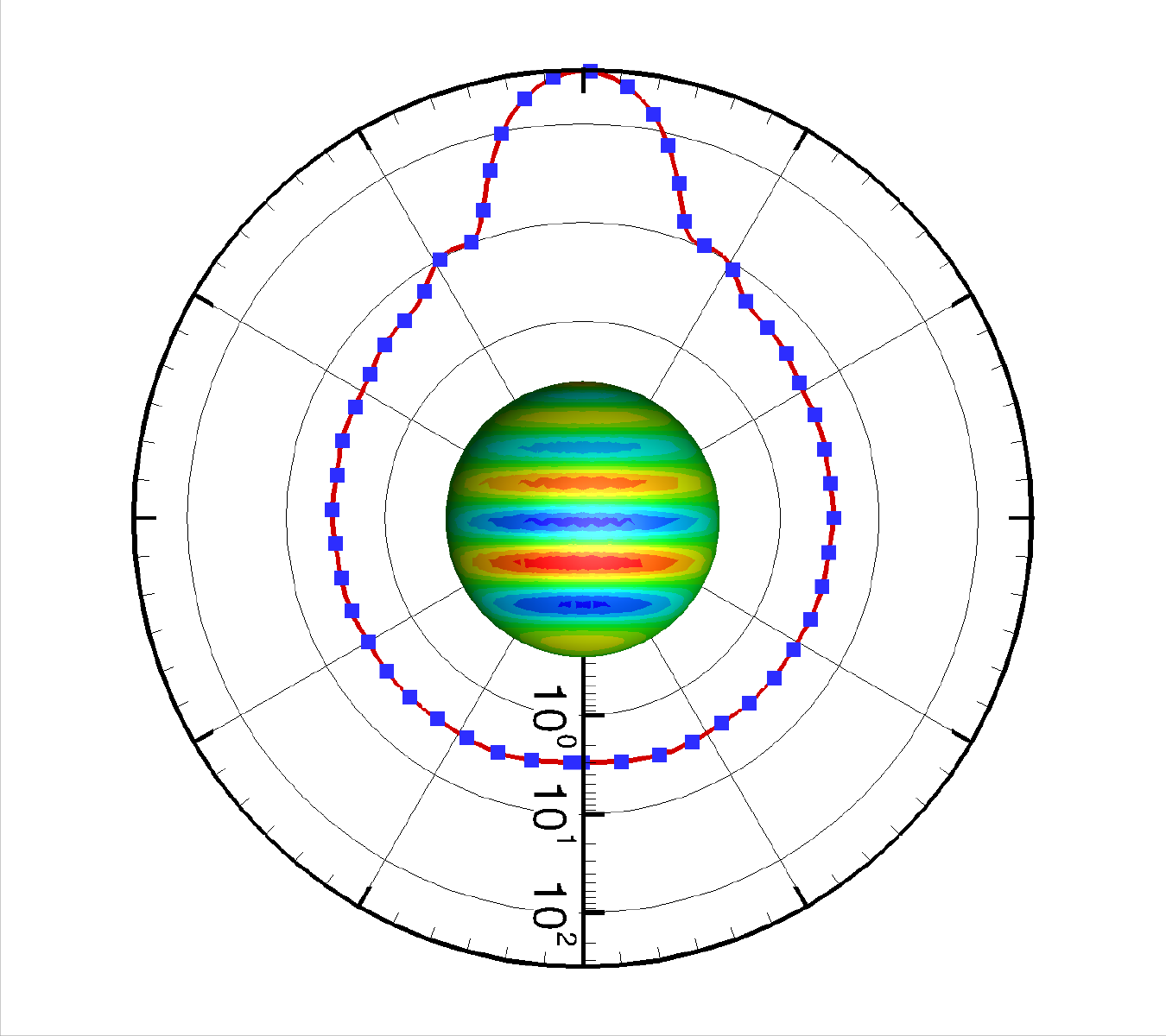} }
  \subfloat[$ka=10, y=0$]{ \includegraphics[width=3.3in]{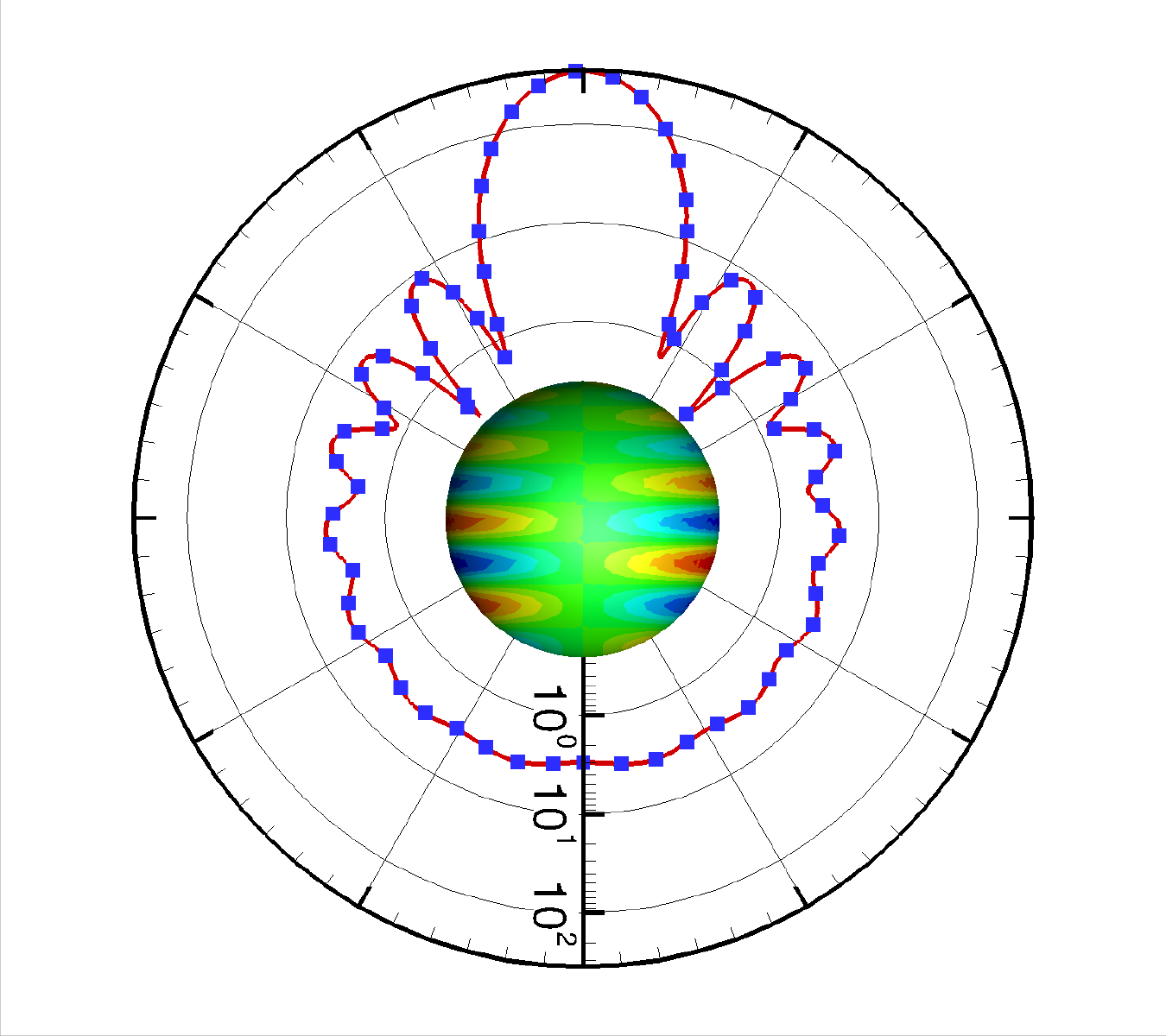} }
\caption{\label{Fig:Mievalid} Comparisons between results from the present field-only formulation (symbols) and from the analytical Mie theory (solid lines). The normal components of the scattered field, $E^s_n$ along the meridian line at $y=0$ on the PEC sphere surface at (a) $ka=1$, using 362 nodes and 180 quadratic elements and (b) $ka=10$, using 1962 nodes and 980 quadratic elements. The induced surface charge density (color sphere) and far field RCS at $r = 20a$ in the planes (c) $x=0$ and (d) $y=0$ for $ka=10$ (color online).} 
\end{figure*}

We show how the solution of (\ref{eq:Helm_E}) and (\ref{eq:Helm_xE}) for the electric field $\boldsymbol{E}$ on the surface of a PEC scatterer can be formulated as a system of linear equations that is the discretized representation of 4 non-singular boundary integral equations  (\ref{eq:NS_BIE}) for the solution of 3 scalar Helmholtz equations for the three components of $\boldsymbol{E}$ and an additional scalar Helmholtz equation for ($\boldsymbol{r} \cdot \boldsymbol{E}$). The total field, $\boldsymbol{E}$, can be written as the sum of the incident and scattered fields: $\boldsymbol{E} = \boldsymbol{E}^i + \boldsymbol{E}^s$. Clearly the known incident field, $\boldsymbol{E}^i$, such as a plane wave, satisfies (\ref{eq:Helm_E}) and (\ref{eq:Helm_xE}), so we only need to solve for the unknown scattered field, $\boldsymbol{E}^s$. On the surface of an object, it is convenient to work in terms of the normal and tangential components of the scattered field: $\boldsymbol{E}^s=\boldsymbol{E}^s_{n}+\boldsymbol{E}^s_{t}$. Since the tangential component of the total field, $\boldsymbol{E}$ must vanish on the surface of a PEC, then the tangential components of the scattered and incident fields must cancel, that is, $\boldsymbol{E}_t \equiv \boldsymbol{E}^s_{t} + \boldsymbol{E}^i_{t} = \boldsymbol{0}$. Thus the components of the scattered field, $\boldsymbol{E}^s = (E^s_x, E^s_y, E^s_z)$ on the surface of a PEC can be expressed in terms of the known tangential components of the incident field, $\boldsymbol{E}^i_t = (E^i_{t,x}, E^i_{t,y}, E^i_{t,z})$, the components of the surface unit normal, $\boldsymbol{n} = (n_x, n_y, n_z)$ with the unknown being the normal component of the scattered field, $E^s_n$ as follows:
\begin{align}\label{eq:ExEn}
 E_{x}^s = E_{n}^s \; n_x - E_{t,x}^{i}
\end{align}\begin{align}\label{eq:EyEn}
 E_{y}^s = E_{n}^s \; n_y - E_{t,y}^{i}
\end{align}
\begin{align}\label{eq:EzEn}
 E_{z}^s = E_{n}^s \; n_z - E_{t,z}^{i}
\end{align}

We discretize the surface, $S$ using quadratic triangular area elements where each element is bounded by 3 nodes on the vertices and 3 nodes on the edge, see Fig. \ref{Fig:quadratic} for a total of $N$ nodes on the surface. The coordinates of a point within each element and the function value at that point are obtained by quadratic interpolation from the values at the nodes using the standard quadratic interpolation function ($\nu \equiv 1 - \xi - \eta$)
\begin{align}
  \phi = \; & \nu(2\nu -1) \; \phi_{1} + \xi(2\xi -1) \; \phi_{2}+\eta(2\eta -1) \; \phi_{3} \nonumber \\
  &  +4 \nu\xi  \; \phi_{4} +4\xi \eta  \; \phi_{5} +4 \eta \nu  \; \phi_{6},
\end{align}
in terms of the local coordinates $(\xi, \eta)$ (see Fig. \ref{Fig:quadratic}).

The solution of  (\ref{eq:Helm_E}) and (\ref{eq:Helm_xE}) for components of the scattered field, $\boldsymbol{E}^s$ and $(\boldsymbol{r} \cdot \boldsymbol{E}^s)$ on the surface are expressed in terms of the values at the $N$ surface nodes. The surface integral solution of these quantities (\ref{eq:NS_BIE}) can be expressed as a system of linear equations in which the elements of the matrices $\cal{H}$ and $\cal{G}$ are the results of integrals over the surface elements involving the unknown $4N$-vector ($E^s_x, E^s_y, E^s_z, \boldsymbol{r}\cdot\boldsymbol{E}^s$). Since the surface integral equation (\ref{eq:NS_BIE}) does not have any singular behavior, these matrix elements can be calculated accurately using standard Gauss quadrature. The linear system can be written as
\begin{align}
  {\cal{H}} \cdot E^s_{x} &= {\cal{G}} \cdot (\partial{E^s_x}/\partial{n}) \label{eq:exbim} \\
  {\cal{H}} \cdot E^s_{y} &= {\cal{G}} \cdot (\partial{E^s_y}/\partial{n}) \label{eq:eybim}\\
  {\cal{H}} \cdot E^s_{z} &= {\cal{G}} \cdot (\partial{E^s_z}/\partial{n}) \label{eq:ezbim}\\
  {\cal{H}} \cdot (\boldsymbol{r}\cdot \boldsymbol{E}^s) &= {\cal{G}} \cdot [\partial{(\boldsymbol{r}\cdot \boldsymbol{E}^s)}/\partial{n}] \label{eq:edotxbim}
\end{align}

For the left hand sides of (\ref{eq:exbim}) to (\ref{eq:ezbim}), we use (\ref{eq:ExEn}) to (\ref{eq:EzEn}) to eliminate the Cartesian components: $E^s_{x}$, $E^s_{y}$ and $E^s_{z}$ in terms of the normal component, $E^s_{n}$, and the tangential component of the known incident field, $\boldsymbol{E}_{t}^{i}$. For Eq. (\ref{eq:edotxbim}), we use Eqs. (\ref{eq:ExEn}) to (\ref{eq:EzEn}) to write
\begin{align}
  \boldsymbol{r}\cdot \boldsymbol{E}^s=(\boldsymbol{r}\cdot \boldsymbol{n})E^s_{n}-(\boldsymbol{r}\cdot \boldsymbol{E}_{t}^{i})
\end{align}
and 
\begin{align}
  \frac{\partial{(\boldsymbol{r}\cdot \boldsymbol{E}^s)}}{\partial{n}}=E^s_{n}+\boldsymbol{r}\cdot \frac{\partial{\boldsymbol{E}^s}}{\partial{n}}
\end{align}
Thus (\ref{eq:exbim}) to (\ref{eq:edotxbim}) can be expressed in terms of the normal component $E^s_{n}$ and the 3 components of the normal derivative $\partial{\boldsymbol{E}^s}/\partial{n}$ of the scattered field as
\begin{align}
  {\cal{H}} \cdot (n_x E^s_n) - {\cal{H}} \cdot E_{t,x}^{i} &= {\cal{G}} \cdot (\partial{E^s_x}/\partial{n})  \\
  {\cal{H}} \cdot (n_y E^s_n) - {\cal{H}} \cdot E_{t,y}^{i} &= {\cal{G}} \cdot (\partial{E^s_y}/\partial{n}) \\
  {\cal{H}} \cdot (n_z E^s_n) - {\cal{H}} \cdot E_{t,z}^{i} &= {\cal{G}} \cdot (\partial{E^s_z}/\partial{n}) \\
  {\cal{H}} \cdot (\boldsymbol{r}\cdot \boldsymbol{n}) E^s_{n}  - {\cal{H}} \cdot (\boldsymbol{r}\cdot \boldsymbol{E}_{t}^{i})&= {\cal{G}} \cdot \left[E^s_{n}+\boldsymbol{r}\cdot \frac{\partial{\boldsymbol{E}^s}}{\partial{n}} \right] 
\end{align}
The above set of equations is a $4N\times 4N$ linear system for the unknown complex $4N$-vectors: $\{\partial{E^s_x}/\partial{n}, \partial{E^s_y}/\partial{n}, \partial{E^s_z}/\partial{n}, E^s_n\}$ on the surface in the final form
\begin{align}
 & \begin{bmatrix}
    -{\cal{G}} & 0 & 0 & {\cal{H}}n_x \\
    0 & -{\cal{G}} & 0 & {\cal{H}}n_y \\
    0 & 0 & -{\cal{G}} & {\cal{H}}n_z \\ 
    -{\cal{G}}x & -{\cal{G}}y & -{\cal{G}}z & {\cal{Y}} \end{bmatrix}
  \left[ \begin{array}{c} \partial{E^s_x}/\partial{n}\\ \partial{E^s_y}/\partial{n} \\ \partial{E^s_z}/\partial{n} \\ E^s_n \end{array} \right] 
 =  \left[ \begin{array}{c}  {\cal{H}}  E_{t,x}^{i}\\ {\cal{H}}  E_{t,y}^{i} \\ {\cal{H}}  E_{t,z}^{i} \\ {\cal{Z}} 
 \end{array} \right].
\end{align}
where ${\cal{Y}} \equiv -{\cal{G}}+{\cal{H}}(\boldsymbol{r}\cdot \boldsymbol{n})$ and ${\cal{Z}} \equiv {\cal{H}} (\boldsymbol{r}\cdot \boldsymbol{E}_{t}^{i})$. This is the linear system to be solved for the surface values of the normal component of the scattered field, $E^s_n$ and the 3 components of normal derivatives ($\partial \boldsymbol{E}^s/ \partial n$).

In a similar way, we can construct the linear system by solving (\ref{eq:Helm_H}) and (\ref{eq:Hn_eq_0}) together with (\ref{eq:Et}) for the tangential components of the $\boldsymbol{E}$ field on the surface. In this case, there are $5N$ unknowns comprising the $2N$ unknowns for the tangential components of $\boldsymbol{H}$ and $3N$ unknowns for the components of ($\partial \boldsymbol{H}^s/ \partial n$).

In contrast to the familiar PMCHWT formulation, the coefficient matrix of our linear systems are well-behaved because of the absence of singularities in our surface integral equations (\ref{eq:NS_BIE}). Values of the surface field on the PEC scatterer - the normal component of $\boldsymbol{E}$ and the tangential components of $\boldsymbol{H}$ are obtained directly. In addition we also obtain the normal derivatives of the fields at the surface. Such quantities are  often sought in surface plasmon applications. In certain EM modeling, the surfaces are assumed to have mathematically sharp corners or edges. For such idealized representations of geometric features, the surface normals and the normal derivatives of surface fields are undefined even though no such difficulties occur with actual physical problems. Thus, a more realistic representation of the details of such geometric features would avoid any unphysical behavior.

\section{Validation and illustrative examples}
We demonstrate the key features and advantages of our field-only formulation with the scattering of an incident plane wave by different PEC objects: (A) a single PEC sphere for which the analytic Mie solution~\cite{vdHulst_1957, Liou_1977} is available for validation; (B) 3 PEC spheres in a triangular configuration in which 2 spheres are nearly touching and (C) a 3D ellipsoid that has aspect ratio 1:3:9. The coupled Helmholtz equations (\ref{eq:Helm_p}) are solved using the non-singular formulation (\ref{eq:NS_BIE}) for the scattered field that are implemented with quadratic surface elements as detailed in the preceding section. Results are designated as:
\begin{itemize}
\item[$1)$]``PEC-E'': ~~~~~if based on (\ref{eq:Helm_E}), (\ref{eq:Helm_xE}) and $\boldsymbol{E}_{t} = \boldsymbol{0}$ on $S$, and
\item[$2)$]``PEC-H'': ~~~~~if based on (\ref{eq:Helm_H}), (\ref{eq:Hn_eq_0}) and (\ref{eq:Et_EQ_0}).
\end{itemize}
We present field quantities on or near the surface of the PEC objects to highlight the utility of our formulation in being able to calculate near fields accurately, in contrast to the PMCHWT formulation. All $\boldsymbol{E}$ field results that follow are obtained with PEC-E, and all  $\boldsymbol{H}$ field results are obtained with PEC-H.  The induced surface electric current density, $\boldsymbol{J}_s$, can be obtained from the magnetic field on $S$: $\boldsymbol{J}_s = \boldsymbol{H} \times \boldsymbol{n}$, and we also check that far field results, such as the radar cross sections can be obtained accurately with our approach. We normalise numerical results for $\boldsymbol{E}$ by the amplitude of the incident field, $|E_0|$, and $\boldsymbol{H}$ is normalized by $k |E_0|/(\omega \mu)$ to ensure all non-dimensional quantities are of comparable magnitude. Comparisons between PEC-E and PEC-H results for the same problem can also be used to quantify the accuracy of the implementations. 

\subsection{Single PEC Sphere - Mie scattering}

Our PEC-E and PEC-H results are checked against the analytic series solution of the Mie problem of the scattering of a linearly polarized incident plane wave by a PEC sphere of radius, $a$~\cite{vdHulst_1957, Liou_1977}. The incident electric field is polarized in the $x$-direction: $\boldsymbol{E}^{i}=(E_{0},0,0)$ and propagates in the $z$-direction: $\boldsymbol{k}=(0,0,k)$. In Fig. \ref{Fig:Mievalid}a and \ref{Fig:Mievalid}b, we show the normal component of the scattered field $E_{n}^{s}$ on the surface of the PEC sphere  along the meridian line in the plane $y=0$ that is calculated by the linear system introduced above. In Fig. \ref{Fig:Mievalid}c and \ref{Fig:Mievalid}d, we show results for the induced surface charge density that is proportional to the normal component of the total electric field and the radar cross section computed from the far field values at $r=20a$. From these, we see excellent agreement between the results calculated by our field-only formulation and the analytical Mie theory.

It is straightforward to show that the resonant modes that arise from our PEC-E or PEC-H solution of a spherical cavity with a PEC boundary~\cite{Harrington_1989} are given by the zeroes of the spherical Bessel functions of the first kind of order $n = 1, 2,...$: $j_n(k_{np}a) = 0$, $p = 1, 2,...$. These are the TE modes~\cite{Harrington_1989} for which the lowest resonant wave number is $k_{11}a = 4.493409$. For example, our numerical solutions are only affected by the resonant solution when $k$ is within 0.1\% of $k_{11}$  using 642 nodes and 320 quadratic elements so the resonant solution is unlikely to affect practical numerical calculations.

\begin{figure}[!t]
\centering{}\includegraphics[width=3.4in]{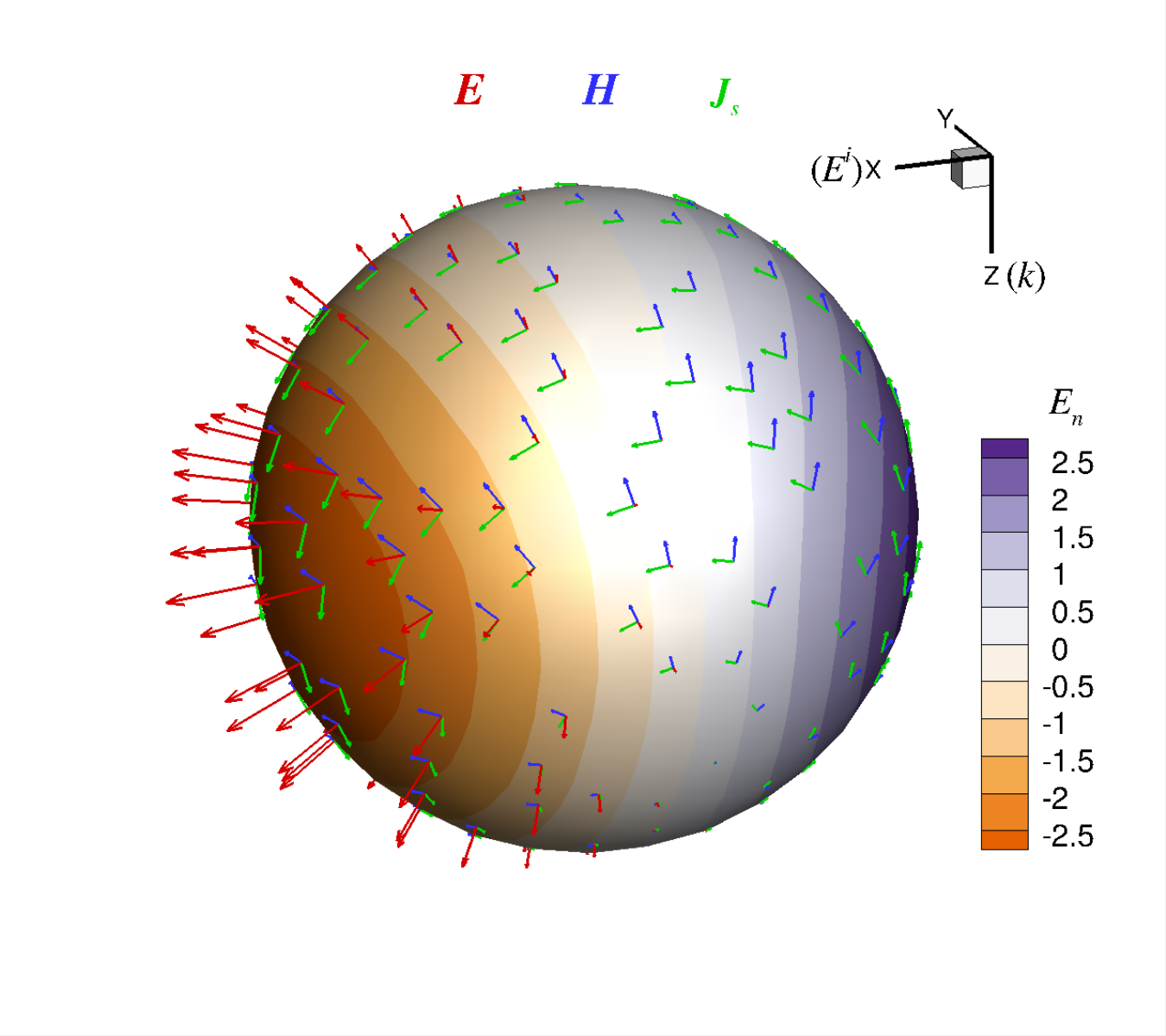} \caption{\label{Fig:Miefield1} Results for the normal component of the total field, $E_n$ (color scale), the total fields $\boldsymbol{E}$ and $\boldsymbol{H}$, and the induced electric surface current density $\boldsymbol{J}_s$ at selected locations on the surface of a perfect conducting sphere of radius, $a$ due to an incident electric field, $\boldsymbol{E}^{i} = (1,0,0) \exp (-jkz)$ with $ka = 1$, obtained using 642 nodes and 320 quadratic elements (color online).}
\end{figure}

In Fig. \ref{Fig:Miefield1}, we show the magnitude of the normal component of the total electric field, $E_n=\boldsymbol{E} \boldsymbol{\cdot} \boldsymbol{n}$ that is proportional to the induced surface charge together with the total electric, $\boldsymbol{E}$, and magnetic, $\boldsymbol{H}$ field vectors as well as the induced surface current density, $\boldsymbol{J}_s$ on the sphere surface at $ka = 1$.

\begin{figure}[!t]
\centering{}\includegraphics[width=3.4in]{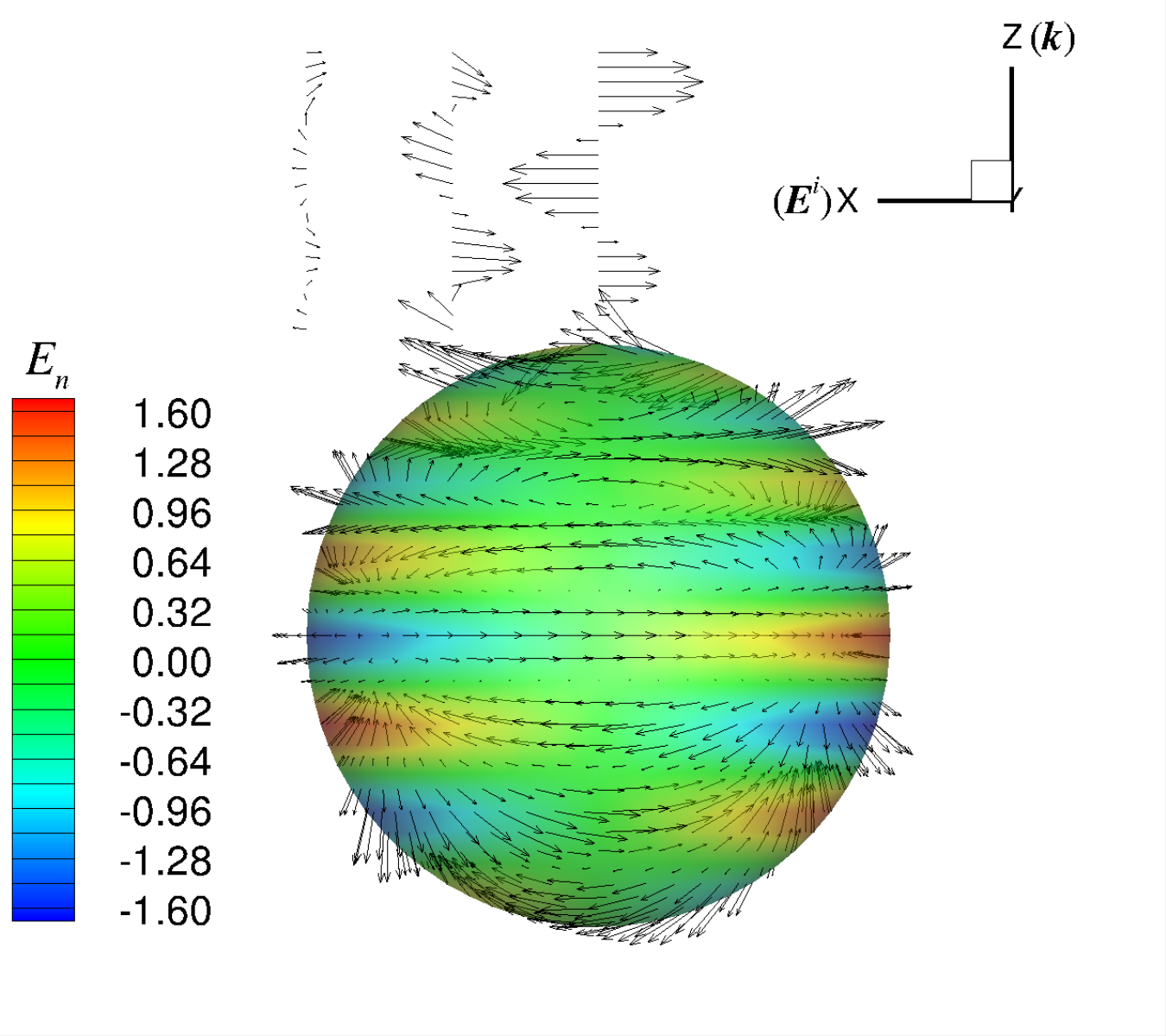} \caption{\label{Fig:Miefield10} Scattered electric field (arrows) on and near the surface of a perfect conducting sphere of radius $a$ and the normal component of the total field $E_n$ (color scale), due to the same incident field as Fig. 1 with $ka = 10$, obtained using 1442 nodes and 720 quadratic elements (color online).}
\end{figure}

In Fig. \ref{Fig:Miefield10}, we show the magnitude of the normal component of the total electric field, $E_n$ at $ka = 10$ as contours together with the scattered electric field on and near the surface. 

\subsection{Three PEC Spheres}
\begin{figure}[!t]
\centering{}\includegraphics[width=3.4in]{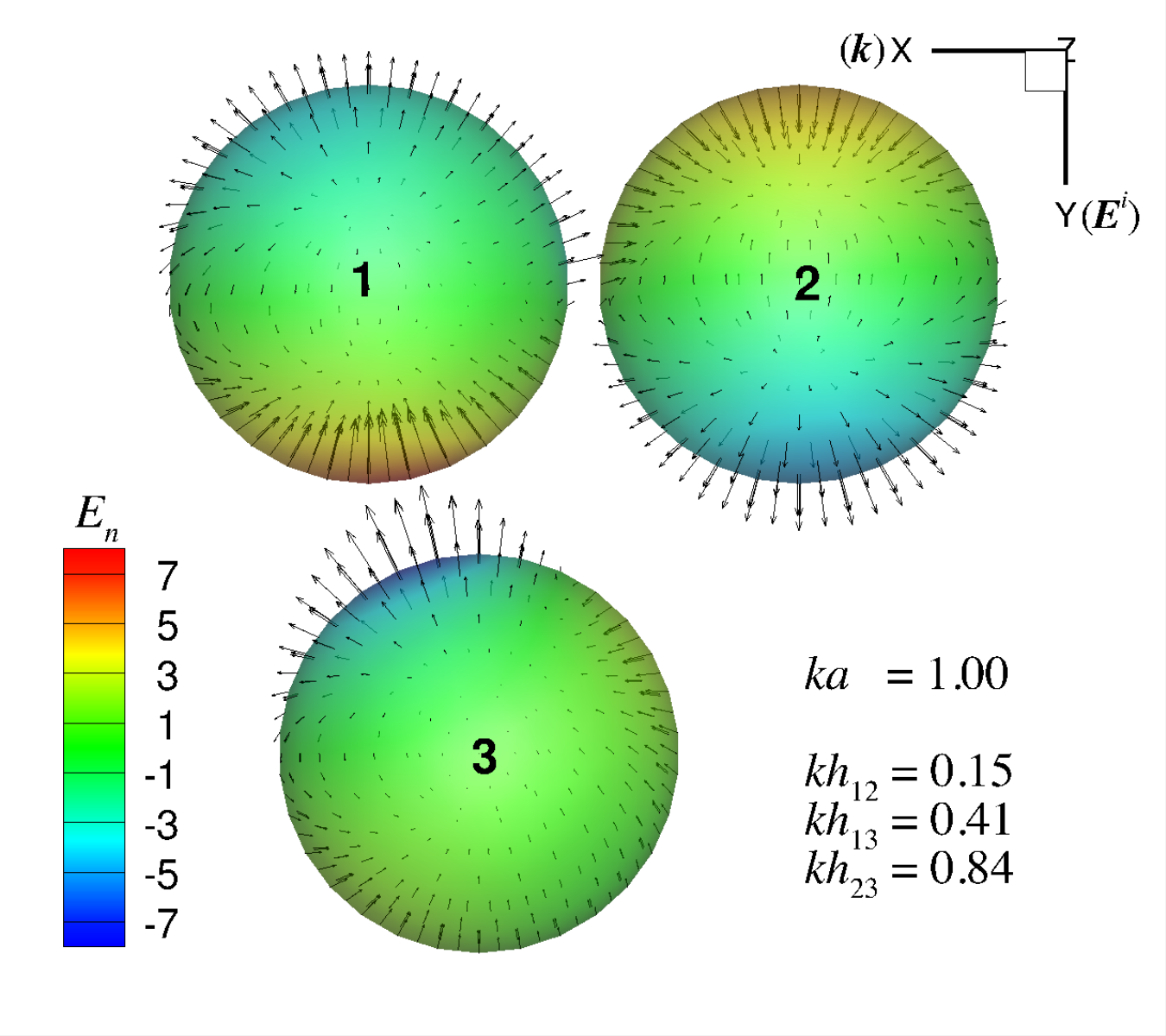} \caption{\label{Fig:3spheres} Scattered electric field $\boldsymbol{E}^s$ (arrows) on the surfaces of 3 identical perfect conducting spheres of radius $a$ and the normal component of the total field $E_n$ (color scale), due to an incident electric field $\boldsymbol{E}^{i} = (0,1,0) \exp (-jkx)$ with $ka = 1$. The distance of closest approach between each pair of spheres $h_{ij}$ is indicated in the figure. The results are obtained using 362 nodes and 180 quadratic elements on each sphere (color online).}
\end{figure}

The absence of singular integrands in our boundary integral solution of our field-only formulation means that closely spaced surfaces will not cause degradation of numerical precision in multiple scattering problems. We consider the scattering of an incident plane wave by 3 identical PEC spheres with $ka = 1$, in a general triangular configuration. The distance of closest approach, $h_{ij}$ between spheres 1, 2 and 3 are $kh_{12}=0.15$, $kh_{13}=0.41$ and $kh_{23}=0.84$. In Fig. \ref{Fig:3spheres} we show the magnitude of the normal component, $E_n$ of the total field and the scattered electric field, $\boldsymbol{E}^{s}$ on the spheres obtained by the PEC-E method.

\subsection{3D PEC Ellipsoid}
\begin{figure}[!t]
\centering{}\includegraphics[width=3.4in]{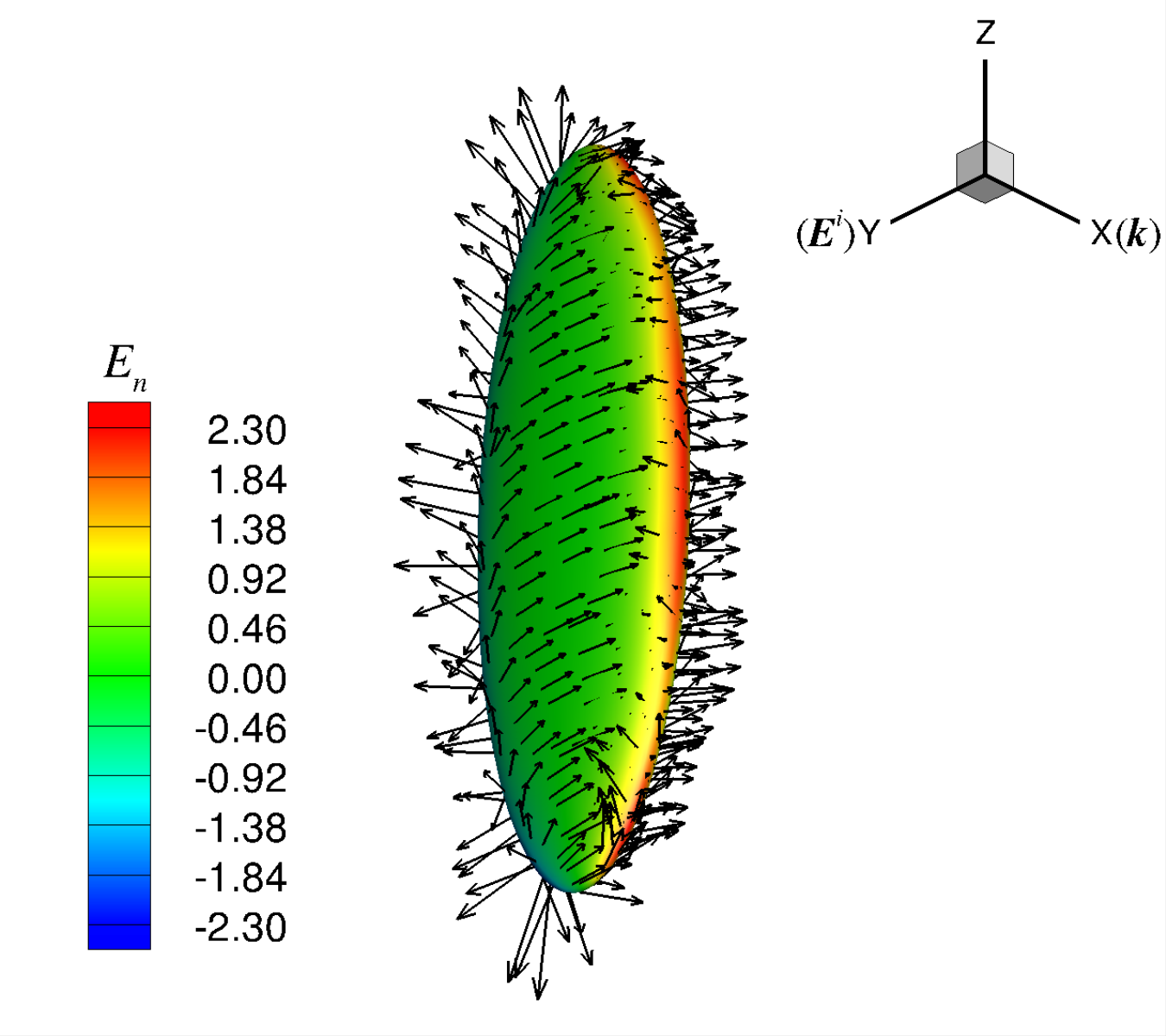} \caption{\label{Fig:ellipsoid} Scattered electric field $\boldsymbol{E}^s$ (arrows) and the normal component of the total field $E_n$ (color scale) on the surface of an ellipsoid with semimajor axes $a, 3a$ and $9a$ due to an incident electric field  $\boldsymbol{E}^{i } = (0,1,0) \exp (-jkx)$ with $ka = 1$, obtained using 2562 nodes and 1280 quadratic elements (color online).}
\end{figure}

To illustrate the capability of our field-only formulation in handling scatterers with a wide range of aspect ratios, we consider the scattering of a plane wave by a 3D PEC ellipsoid whose surface is given by: $(x/a)^2 + (y/3a)^2 + (z/9a)^2 = 1$, at $ka = 1$. The magnitude of the normal component, $E_n$ of the total field and the scattered electric field, $\boldsymbol{E}^{s}$ on the ellipsoid are shown in Fig. \ref{Fig:ellipsoid}.

\section{Conclusion}

We have developed a formulation of electromagnetics in the frequency domain that only involves the electric field, $\boldsymbol{E}$ or the magnetic field, $\boldsymbol{H}$. This is a simpler alternative to the established PMCHWT approach.  Our formulation only involves solving scalar Helmholtz equations for the components of $\boldsymbol{E}$ or $\boldsymbol{H}$ and for the scalar functions $(\boldsymbol{r} \cdot\boldsymbol{E})$ or $(\boldsymbol{r} \cdot\boldsymbol{H})$.  The PEC-E formulation gives rise to $4N$ unknowns as only the normal component of $\boldsymbol{E}$ is unknown whereas with the PEC-H formulation, both tangential components of $\boldsymbol{H}$ are unknown and thus it gives rise to $5N$ unknowns. Indeed the ability to obtain the same numerical solution using the PEC-E $4N$ system and using the PEC-H $5N$ system provides an internal check of the consistency of our theoretical formulation and accuracy of the numerical implementation.

A non-singular boundary integral method~\cite{Sun_2015} is used to solve the Helmholtz equation that is easy to implement and affords much higher precision than conventional numerical methods as quadratic elements can be readily employed. Consequently, it is no longer necessary to work with electric and magnetic surface currents as intermediate quantities as required in the PMCHWT formulation. However, if required, surface currents can be readily found by post-processing. This affords considerable simplification in implementation compared to that of surface current basis functions such as the popular RWG scheme. The immediate availability of surface field values without further post processing may be desirable in studies of surface enhanced Raman effects as well as in photonic and plasmonic applications. Thus relative to the current-based surface integral formulation that requires further post processing by taking numerical derivatives of the surface current to obtain the surface fields, the present approach yields the surface fields directly at the expense of working with a larger number of degrees of freedom, but this is compensated by the ability to use quadratic elements that can furnish higher precision with fewer unknowns.  The balance of this trade-off may be a topic for future evaluation.

The absence of singularities in the integral equation formulation of the Helmholtz equations means that surface integrals can be calculated accurately using standard quadrature. The removal of the singularity has no adverse effect on the condition number of the linear system~\cite{Sun_2015}. Furthermore, problems that have boundaries that are close together will no longer suffer degradation of numerical stability and precision \cite{Sun_2015}. In all our examples, only a very modest number of nodes are needed. The solution of the integral equations can be accelerated to be a $O(N \log N)$ problem using fast Fourier transform and fast multipole methods~\cite{Cao_2015}.

Since the present formulation works directly with field values on the surface there remains the open question of modeling boundaries that have mathematically sharp edges and corners. At such idealized geometric singularities, the surface field values are physically not defined. Therefore, more investigation is needed for the application of this formulation to non-smooth surfaces.

\section*{Acknowledgment}
We thank William Stewart and Albert Zhang for their assistance in implementing the Mie theory in \textit{Mathematica} and \textit{Matlab}. This work is supported in part by the Australian Research Council through a Discovery Early Career Researcher Award to QS and a Discovery Project Grant to DYCC.


%

%

%
%
%
%
%
%
%
%
%
%
%
%
%
\newpage
%
%
%
%
%
%




\end{document}